\documentclass[%
 aip,
 amsmath,amssymb,
 reprint,%
]{revtex4-1}

\usepackage{graphicx}
\usepackage{dcolumn}
\usepackage{bm}
\usepackage{caption}
\usepackage{subcaption}
\usepackage{graphicx}
\usepackage[version=4]{mhchem}

\usepackage[utf8]{inputenc}
\usepackage[T1]{fontenc}
\usepackage{mathptmx}
\usepackage{xcolor}
\usepackage{natbib}
\begin{document}

\preprint{AIP/123-QED}

\title[]{Transformation of Analog to Digital Resistive Switching in Cu Implanted ITO/NiO/Ag Device for Neuromorphic Applications 
 }

\author {Sourav Bhakta}
\affiliation { 
School of Physical Sciences, National Institute of Science Education and Research (NISER) Bhubaneswar, an OCC of HBNI, Jatni-752050, Odisha, India 
}%

\author{Pratap K. Sahoo}
 \email{pratap.sahoo@niser.ac.in}
\affiliation { 
School of Physical Sciences, National Institute of Science Education and Research (NISER) Bhubaneswar, an OCC of HBNI, Jatni-752050, Odisha, India 
}%
\affiliation { 
Center for Interdisciplinary Sciences (CIS), NISER Bhubaneswar, HBNI, Jatni-752050, Odisha, India
}%

\begin{abstract}
Both analog and digital resistive switching are essential components in the neuromorphic computing system. This work reports the influence of Cu ions for the transformation of analog to digital resistive switching in Indium–Tin-Oxide(ITO)/NiO/Ag memristor devices. The undoped and low-concentration Cu doping illustrates the analog switching, whereas higher doping demonstrates the digital characteristics. At higher bias voltage, the Schottky barrier ($\phi_B$) is developed at both ITO/NiO and NiO/Ag interfaces. The increasing and decreasing of current conduction with the escalating number of cycles for both the polarity in undoped and low doped is elucidated by the electrode-dominated mechanism in terms of reduction and enhancement of Schottky barrier height at the interface, respectively. The digital switching characteristic due to the formation and rupturing of the vacancy filament at higher doped sample is induced due to the boosting of vacancies above the critical amount using ion implantation. The synergic effect of current conduction due to local Cu migration and oxygen vacancies can be utilized as a learning and forgetting process for neuromorphic applications. 


\end{abstract}

\maketitle

\section{\label{sec:level1}INTRODUCTION}

The next generation technological advancement and complex computation capabilities demand high data processing in miniaturized storage devices. Hence, intense research is in progress for an alternative way to the conventional CMOS-based technology to overcome the shortcomings of Si-based flash memories with the physical and technological limitations of building an advanced memory device over the past few decades. Resistive random access memory (RRAM) exhibits the potential interest to replace these limitations due to its simple structure, rapid operation, high-density integration, clear switching events, good reversibility, neuromorphic computing, and 3D staking compatibility \cite{kim2011nanofilamentary,waser2007nanoionics,baek2004highly,sun2015thermal,markovic2020physics}. Among various binary metal oxides such as TiO$_2$ \cite{cao2009structural}, ZnO \cite{chen2012resistive}, HfO$_2$ \cite{privitera2013microscopy}, CeO$_2$ \cite{younis2013interface}, Cu$_x$O \cite{dong2007reproducible}, NiO is one of the suitable candidates due to having simple cubic rack salt structure and intrinsic wide gap (3.70 eV) for RRAM applications to fulfill the above criteria \cite{seo2004reproducible}. The digital RRAM stored information by switching resistance abruptly in the oxide layer from a high resistance state (HRS) to a low resistance state (LRS) and vice versa.
On the other hand, analog switching shows a gradual change in conductance with the bias, which has drawn enormous interest among researchers due to its intrinsic analogy to biological synapses for implementing brain-inspired neuromorphic computing applications. Such analog memory analogous to artificial synapses can be utilized to understand the highly efficient bio-mimicking computing system \cite{del2018challenges}. Various approaches, such as doping, interface engineering, implantation, etc., have been adopted for the betterment of the memristor parameter to realize the high-performance device \cite{shi2021review, deng2014effects}. Mitra et al. \cite{mitra2014study} showed the anticlockwise bipolar resistive switching in chemically synthesized ITO/NiO/Ag heterojunction. The report explains the switching mechanism in terms of rupture and recovery of conducting filament in the oxide layer incorporated with the migration of oxygen ions and interfacial effects. The effect of Schottky contact of ITO/NiO for the variation of resistive switching is reported by No et al. \cite{no2013effect}. Electrode-dependent digital and analog switching in NiO-based memristors is investigated by Swathi et al. \cite{swathi2022digital}. Saleh et al. \cite{saleh2010effect} explained the impact of Ni (260 and 75 KeV) and O (80 and 25 KeV) ion implantation in the variation of resistive switching of Pt/NiO/Pt devices. Elliman showed the fabrication of Ta$_2$O$_5$/TaO$_x$ heterostructure by O ion implantation beyond the conventional film deposition techniques to observe the application of implantation for improving the non-volatile RRAM device. Ning et al. \cite{deng2014effects} investigated the change in resistive switching characteristics in HfO$_2$ by implanting 30 KeV Al, Cu, and N ions at a dose of $1\times10^{14}$ ions/cm$^2$. The report investigated a remarkable improvement in device-to-device uniformity by only Cu ion implantation. Yen et al. \cite{yen2021improved} observed high stability and excellent retention in 10 KeV As$^+$ ion implanted SiN$_x$ memristor compared to the nonimplanted device. Hardly reports were found showing both analog and digital switching in NiO. However, the analog to digital resistive switching using Cu-ion implantation in ITO/NiO/Ag devices has not been explored.

In this report, we investigated the impact of 100 KeV Cu ion implantation in ITO/NiO/Ag memristor to observe the transformation from analog to digital switching. The introduction of vacancy defects in NiO induces digital switching in high-dose Cu doping, which is found to be bulk-dependent; in contrast, analog switching is electrode-dominated (Pristine and low Cu dose sample). The decrease and increase of the Schottky barrier have been observed for pristine and low-dose samples. This enhances and reduces the current conduction with the number of cycles in pristine and low-dose samples, respectively. The current conduction behavior can be combined together for neuromorphic applications. The digital switching is activated in high-dose samples due to the boosting of vacancies above the critical amount in the NiO matrix by ion implantation.  

\section{\label{sec:level2}Experimental Procedure}
The ITO substrate was brought from MTI Corporation. The substrate was cleaned with acetone, isopropyl alcohol, and de-ionized water, respectively, after sonicating with each of them for 5 minutes. The cleaned ITO substrate was mounted to a self-designed Radio-Frequency (RF) sputtering system perpendicular to the NiO target (2-inch diameter, $99.99\%$ purity). The substrate-to-target distance was kept fixed at 6.5 cm during the deposition. The sputtering parameters, such as RF power, Ar gas flow rate, base pressure, and deposition pressure, were 100 W, 15 sccm, 6.77 $\times 10^{-6}$ mbar, and 5 $\times 10^{-3}$ mbar, respectively. The as-grown samples were annealed using the plasma-induced chemical vapor deposition (PECVD) system in a vacuum of 5.63 $\times 10^{-2}$ mbar. The annealed NiO sample (pristine) was implanted using 100 keV Cu ions at room temperature, scanning over 10 $\times$ 10 mm area of the sample surface using the electrostatic scanner for uniform implantation. The sample implanted with ion fluences $5\times 10^{15}$ and $2\times 10^{16}$ ions/cm$^2$ are denoted as sample B and sample C, respectively. Low ion flux at $2\times 10^{12}$ ions/cm$^2$ was maintained to avoid beam heating. The range of 100 keV Cu in NiO is 80 nm, calculated from Stopping and Range of Ions in Matters (SRIM) simulations \cite{ziegler2010srim}. The simulation shows that nuclear energy loss (2286 KeV/$\mu m$) dominates over the electronic energy loss 203.8 KeV/$\mu m$, more than 11 times. Hence, the incident ion can displace the host atoms from their lattice site through elastic collision. This yields the creation of vacancy defects in the system. The phase identification of pristine and implanted NiO samples was carried out using the Rigaku Smartlab X-Ray diffractometer with Cu $K_{\alpha}$ source ($\lambda = 1.5418 \AA$). The thickness and surface morphology of the samples are investigated using cross-sectional Field Emission Scanning Electron Microscopy (FESEM) and Atomic Force Microscopy (AFM) images. The top electrode on the NiO sample was accomplished by conducting Ag paste. The I-V measurements of the ITO/NiO/Ag memristor devices were performed with a voltage sweep step of 40 mV using a two-terminal Keithley 2450 source meter.

\section{\label{sec:level4}Results and Discussions}

\begin{figure}[htb!]
\includegraphics [width=9.5cm]{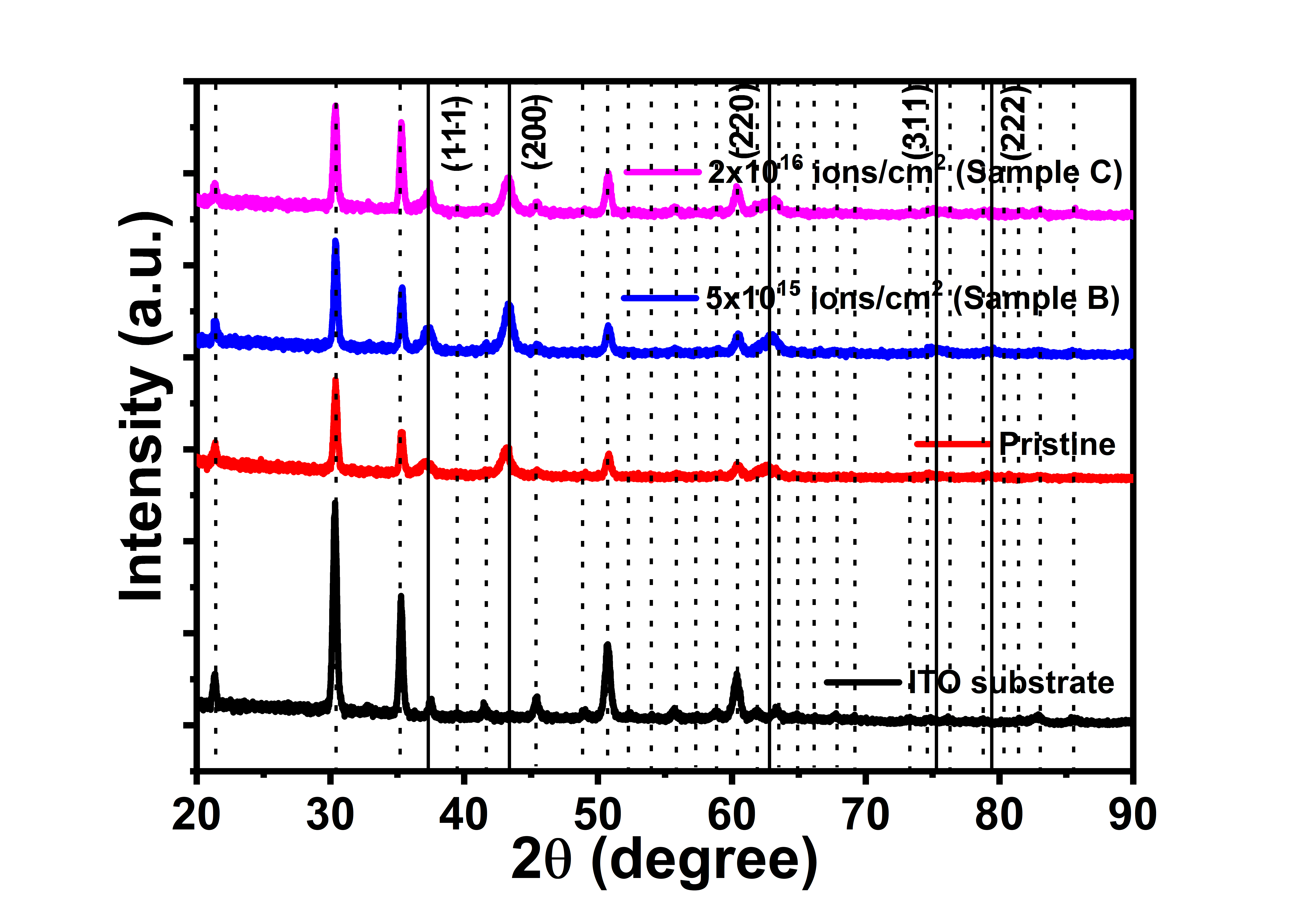}
\caption{\label{fig:picture_1} The X-ray diffraction spectra of ITO substrate, pristine, sample B, and sample C.}
\label{1}
\end{figure}

The X-ray diffraction (XRD) spectra of the ITO substrate, pristine, sample B, and sample C are shown in Fig. \ref{1}. The dashed line in Fig. \ref{1} corresponds to the peak position of the ITO substrate. The solid line corresponds to the NiO peak position. The NiO peaks are observed at 37.31, 43.30, 62.73, 75.37, and 79.40$^{\circ}$ for all the samples. These peaks agree well with the JCPDS card number 780429, which identifies the space group of Fm$\bar 3$m with the cubic phase of NiO. The XRD spectra confirm that the grown NiO film is polycrystalline in nature.

\begin{figure}[htb!]
\includegraphics [width=8cm]{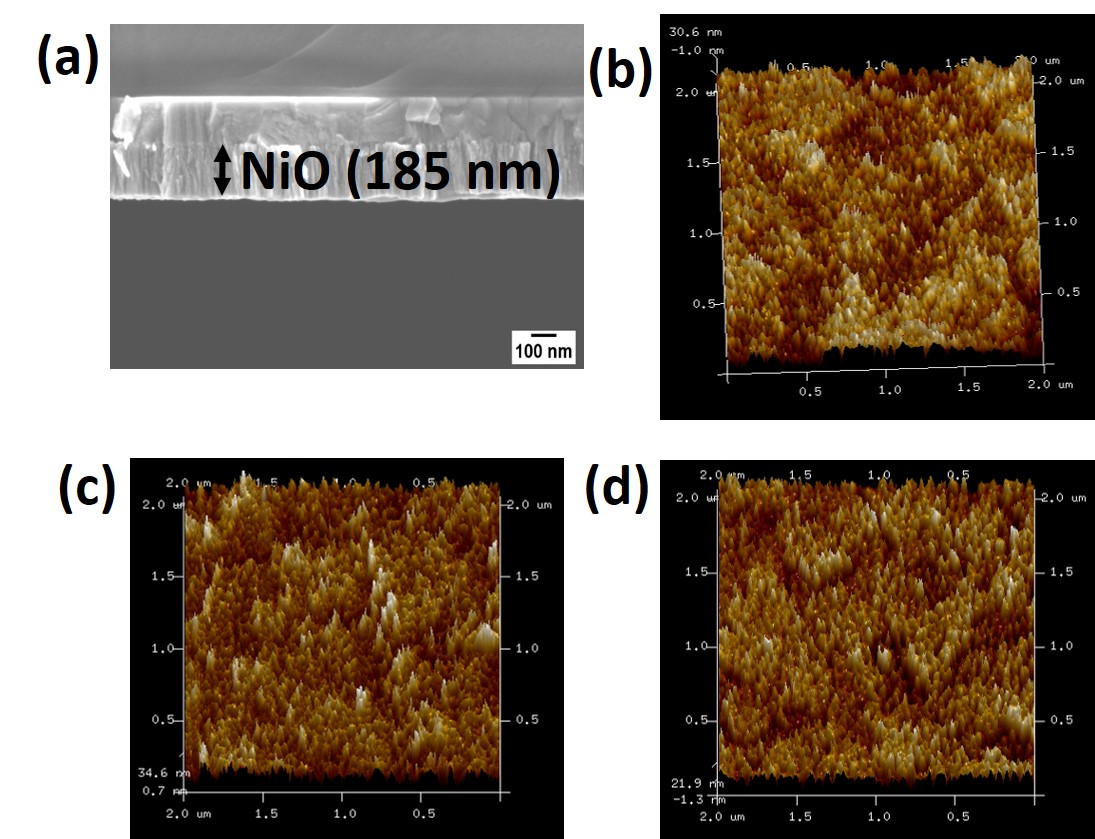}
\caption{\label{fig:picture_1} (a) Crossectional FESEM images of the pristine sample. The AFM images of the (b) pristine, (c) sample B, and (d) sample C. }
\label{1a}
\end{figure}

The thickness of the as-grown film is evaluated using the crossectional FESEM images. Fig. \ref{1a}(a) shows that the film thickness of the pristine sample is found to be 185 nm. The surface morphology of the pristine and implanted samples (sample B and sample C) are examined by AFM images, as shown in Fig. \ref{1a}(b-d), respectively. Pristine sample shows the continuous growth of the NiO thin films. The surface of the implanted samples doesn't show a drastic variation compared to the pristine samples, rather than the roughness. The Root Mean Squared (RMS) surface roughness of the pristine is 8.82 nm, and it increases to 9.02 and 9.38 nm in samples B and C, respectively, due to Cu ion implantation.

\begin{figure*}[htb!]
\includegraphics [width=15cm]{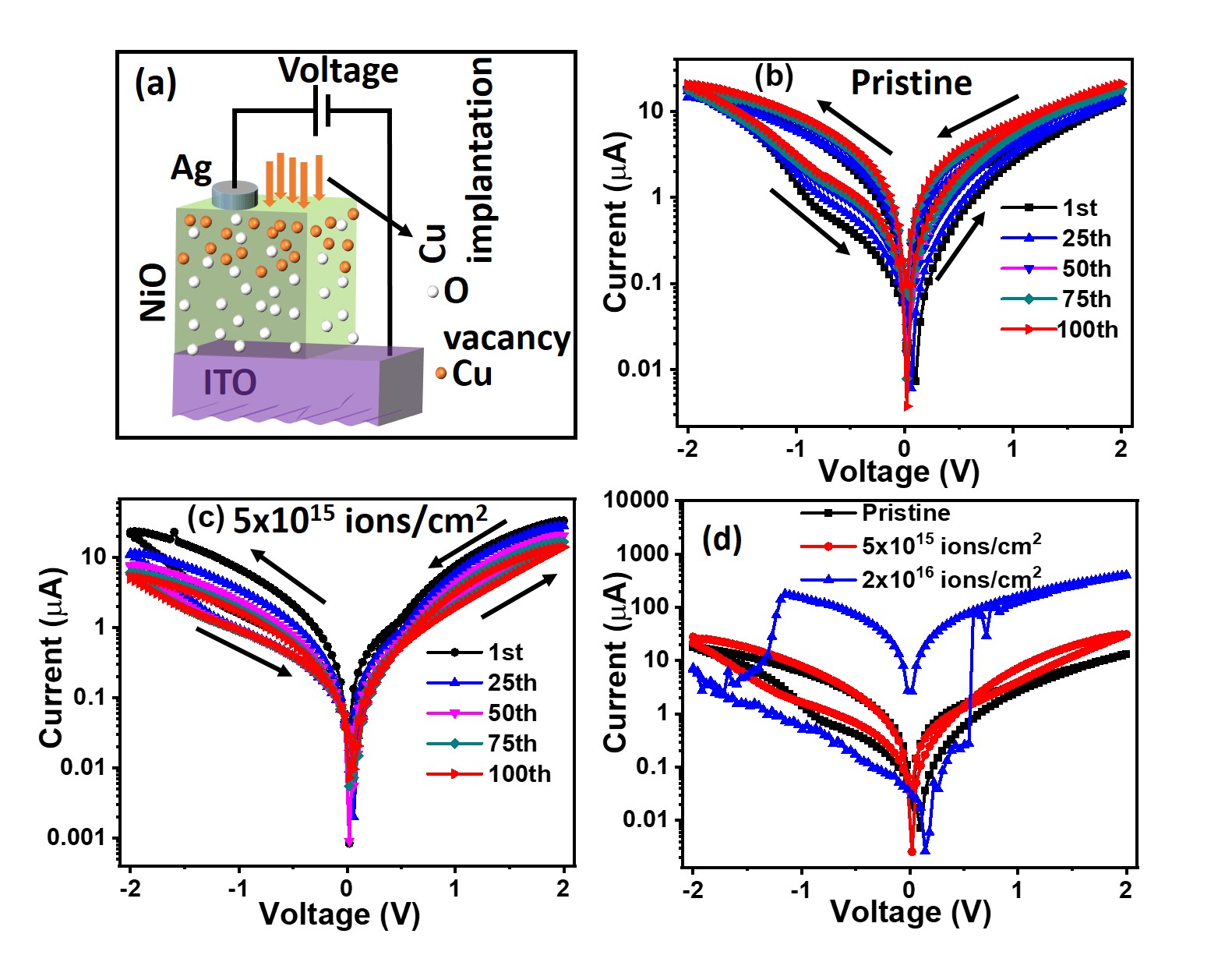}
\caption{\label{fig:picture_1} (a) Schematic of ITO/NiO/Ag RRAM device for I-V measurements. The current-voltage curve of the 1st, 25th, 50th, 75th, and 100th cycles of (b) pristine and (c) sample B. (d) The variation of the current (1st cycle) with an applied bias voltage of pristine and implanted samples with increased ion fluences. }
\label{3b}
\end{figure*}

\begin{figure}[htb!]
\includegraphics [width=8cm]{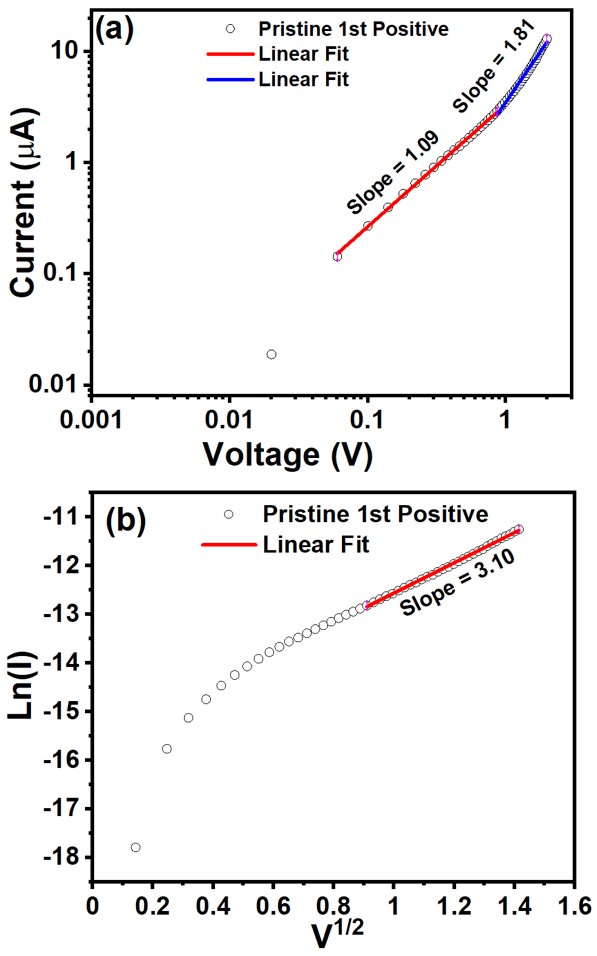}
\caption{\label{fig:picture_1} (a) The double logarithmic I-V plot and (b) ln(I) vs $V^{1/2}$ plot with liner fitting for first positive voltage scan of pristine. }
\label{3a}
\end{figure}

The schematic of the NiO-based ITO/NiO/Ag memristor device is shown in Fig. \ref{3b}(a). The bias voltage is stressed between the top Ag electrode and the bottom ITO substrate. Fig \ref{3b}(b) and \ref{3b}(c) shows the variation of current with applied voltage of NiO memristor of pristine and sample B of the 1st, 25th, 50th, 75th, and 100th cycles for positive and negative bias sweeping. The I-V curve of the 1st cycle with the variation of ion fluences is demonstrated in Fig. \ref{3b}(d). An abrupt change of current at a particular voltage signifies the digital switching in sample C, which will be discussed later. In contrast, pristine and sample B show the bipolar analog switching in the -2 to +2 voltage sweep range. The small-scale asymmetry is observed in pristine. This indicates the formation of the Schottky barrier at the NiO/Ag interface, with NiO being a p-type semiconductor and Ag an n-type metal. The ITO/NiO junction also forms the Schottky barrier, as reported by Swathi and No et al. \cite{swathi2022digital,no2013effect}. The Schottky barrier at both interfaces is crucial for analog switching. The anticlockwise hysteresis is observed for both the samples in both positive (0 V $\rightarrow$ 2 V $\rightarrow$ 0 V) and negative voltage sweep regions (0 V $\rightarrow$ -2 V $\rightarrow$ 0 V). In pristine, the current increases with increasing cycles at both the positive and negative bias sweeping region. In contrast, the opposite scenario is observed in sample B. The decrease of the current in sample B may be attributed to the defect formation in the system. Cu implantation creates a lot of vacancies, interstitial, and substitutional defect states in the film, and the damage increases with ion fluences. When these defect states are stressed by sufficient fields for a long period, the carrier's movement may be disrupted. This can yield a change in the Schottky barrier height and Fermi level, which is responsible for decreasing the current with the consecutive increase in the number of cycles. The negative voltage sweep exhibits a bigger hysteresis compared to the positive voltage sweep. The increase and decrease in conductance in pristine and sample B can be useful for biological synapses \cite{wang2012synaptic}. In biological systems, presynaptic and postsynaptic neurons communicate themselves through the synaptic node. The variation in the concentration of the synapses plays a key role in the brain's memory and learning process \cite{yang2013synaptic}. The short-term memory (STM) followed by long-term memory (LTM) is one of the basis of human memory loss, according to Psychological studies \cite{wixted1991form}. The frequent stimulation of the biological system can transform the STM into LTM \cite{mcgaugh2000memory}. In this work, the increasing or decreasing current of the NiO-based memristor can be modulated gradually with the number of cycles by taking the combination of pristine and sample B. Hence, using this analog resistive switching of NiO, synaptic learning and forgetting characteristics can be stimulated and employed in neuromorphic systems. The increasing current with consecutive pulses in pristine can be assigned as a learning process, and the decreasing current in sample B is analogous to the forgetting process. Now, to investigate the mechanism of analog switching at the interface, we fitted the double logarithmic I-V curve of the 1st positive cycle linearly. Fig. \ref{3a}(a) and \ref{3a}(b) show the linear fitting of double logarithmic I-V curve and ln(I) vs $V^{1/2}$ plot (Schottky plot) of the 1st positive cycle for pristine, respectively. The Ohmic condition is satisfied under the applied low bias voltage (0 V $\rightarrow$ 0.9 V), as the curve is fitted well linearly with a slope of 1.09, as seen in Fig. \ref{3a}(a). But, under a high bias voltage, the deviation of slope from 1 indicates other conductive mechanisms, such as Schottky emission \cite{biju2011resistive}, space charge limited conduction \cite{won2009transparent}, Fowler Nordheim tunneling \cite{chang2013analog}, and Poole Frenkel emission \cite{park2016flexible} that can dominate over Ohmic conduction. The good linear fitting observed in Fig. \ref{3a}(b) at a high bias voltage region suggests the formation of Schottky contact at both interfaces. The good linear fitting of the Schottky plot at the high bias voltage region and the asymmetry characteristic signifies the domination of Schottky emission compared to the other conduction mechanism. We used the following Richardson-Schottky equation to understand the I-V characteristics \cite{zhong2014nonpolar}:

\begin{equation}\label{schott}
    I = AB^*T^2 exp[{-q\phi_{B} + \frac{q}{kT}\sqrt{\frac{qE}{4\pi \epsilon_i}}}]
\end{equation}

Where, A = Conduction area, $B^*$ = Effective Richardson constant ($\frac{4 \pi q m_n^*k_0^2}{h^3}$), T = Absolute Temperature, k = Boltzmann constant, q = electric charge, E = $\frac{V}{d}$, d = NiO film thickness, $\phi_B$ = Schottky barrier, $\epsilon_i$ Dielectric constant of NiO.

After taking logarithm on both sides, the equation \ref{schott} reflects as:

\begin{equation}
    ln(I) = ln (AB^*T^2) - \frac{q\phi_B}{kT} + \frac{q}{kT}\sqrt{\frac{q}{4\pi\epsilon_id}}\sqrt{V}
\end{equation}

\begin{figure}[htb!]
\includegraphics [width=9cm]{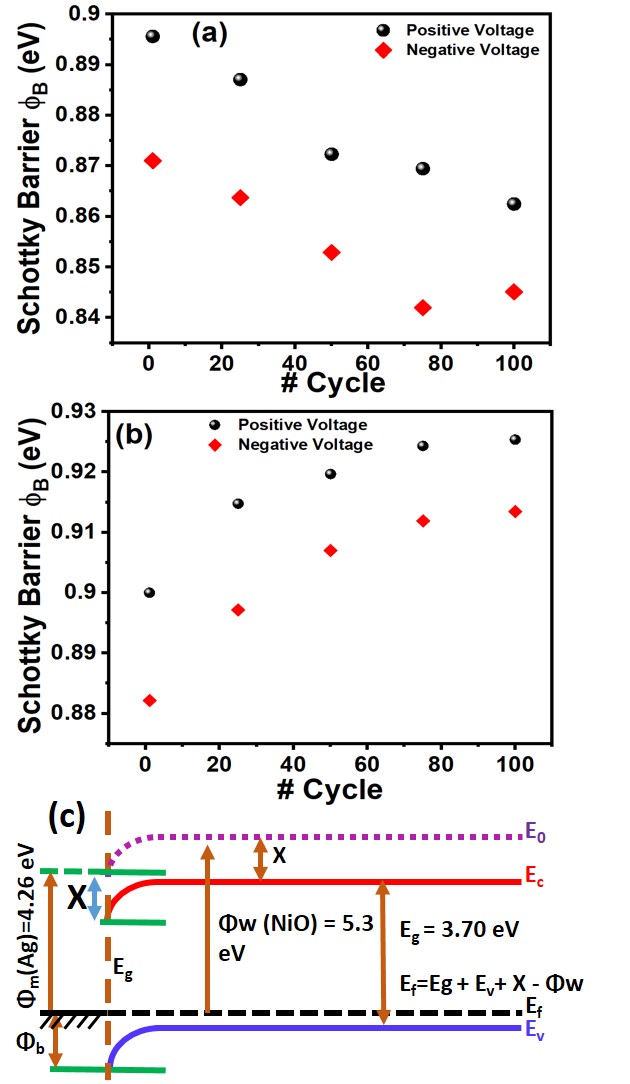}
\caption{\label{fig:picture_1} The variation of Schottky barrier with the increase of the number of cycles in (a) pristine, (b) sample B for positive and negative voltage scan. (c) The schematic of Schottky contact at NiO/Ag interface.  }
\label{4}
\end{figure}

The intercept value at the ln (I) axis after linear fitting of the Schottky plot in positive and negative bias at higher field region (Fig. \ref{3a}(b)) (0.9 V $\rightarrow$ 2 V) provides the Schottky barrier height. In this case, we used the Richardson constant as $B^*$ = 119.8 K/$K^2.cm^2$ by considering $m_n^*$ = 1.0 and $m_0$ = $9.1 \times 10^{-31}$ Kg. The variation of the Schottky barrier with the number of cycles for positive and negative voltage sweep of pristine and sample B is shown in Fig. \ref{4}(a) and \ref{4}(b), respectively. The energy band diagram of Schottky contact is illustrated in Fig. \ref{4}(c). In pristine, the Schottky barrier decreases with the increase of the number of cycles for both positive and negative bias. The obstruction for the carrier reduces due to the decrease in the Schottky barrier at the interface. Hence, the reduction of $\phi_B$ with increasing cycle increases the conduction in pristine for both positive and negative voltage sweep. In contrast, the decrease of the current with the number of cycles in sample B is due to the increase of the Schottky barrier at the interface for both positive and negative voltage sweep. The increase in the barrier height may be attributed to the creation of defects by Cu ion implantation in the NiO thin films. The carriers need more energy to overcome the barrier height, which decreases the current conduction in sample B. Hence, the prevailing conduction mechanism in analog RRAM devices is governed by the interface-dominant Schottky effect rather than filament formation.

\begin{figure}[htb!]
\includegraphics [width=8.5cm]{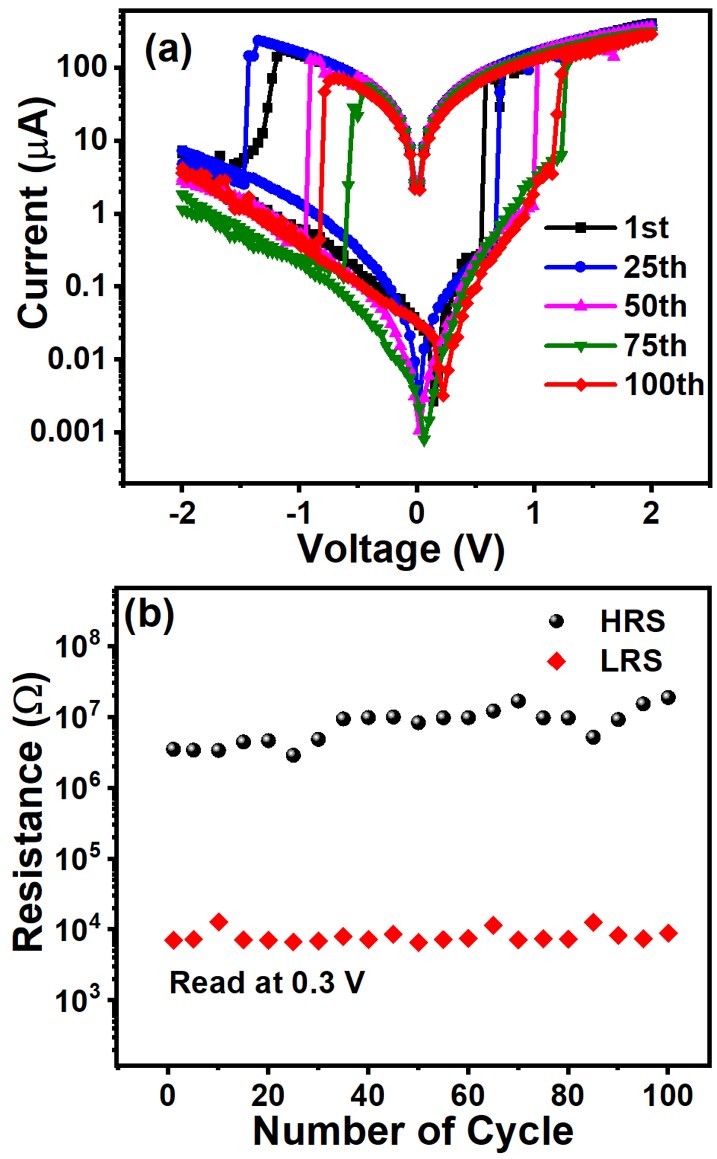}
\caption{\label{fig:picture_1} (a) The current-voltage curve of the 1st, 25th, 50th, 75th, and 100th cycles of sample C, (b) The LRS and HRS state with an increasing number of cycles for the endurance test of sample C. The HRS and LRS was read at 0.3 V. }
\label{retain1}
\end{figure}

\begin{figure*}[htb!]
\includegraphics [width=16cm]{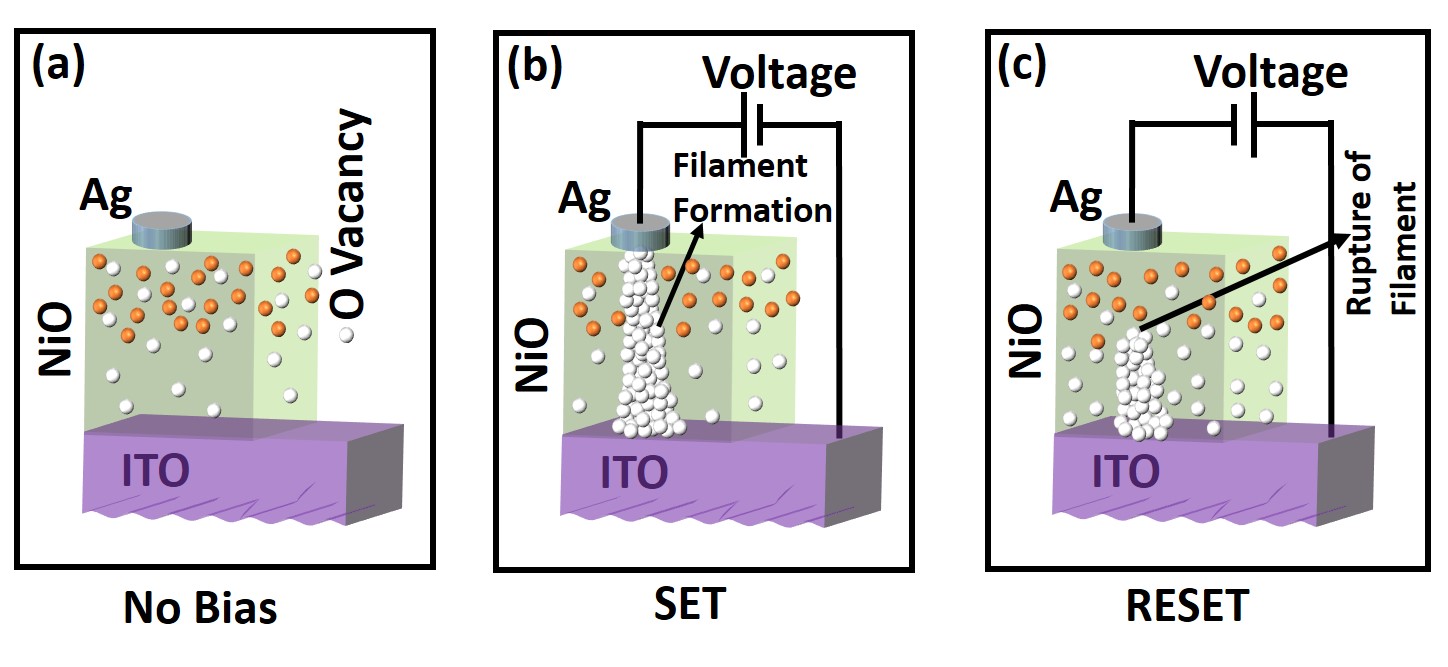}
\caption{\label{fig:picture_1} The schematic of the proposed filamentary model for the (a) zero bias, (b) SET, and (c) RESET process in digital resistive switching of sample C.}
\label{5}
\end{figure*}

Sample C shows the digital bipolar resistive switching in the same voltage sweeping range (-2 V $\rightarrow$ 0 V $\rightarrow$2 V and 2 V $\rightarrow$ 0 V $\rightarrow$ -2 V), as shown in Fig. \ref{retain1}(a). An electroforming process was observed at the lower voltage region due to the presence of high vacancy defect concentration in the sample, induced at high Cu ion fluences. The higher fluence sample is transformed into a digital RRAM device through the electroforming process, whereas the pristine and lower fluence samples (sample B) show the signature of analog switching. The current jumps at 0.54 V for the first cycle and reaches the LRS state, defining it as a SET process. The device moves from LRS to the HRS at -1.15 V by dropping the current abruptly, defining the RESET process. The SET and RESET cycles of the memristor (sample C) are clearly observed in Fig. \ref{retain1}(a). The SET voltage increases with increasing the number of cycles, while the RESET voltage reduces for the higher one with respect to the initial cycle. The hysteresis developed drastically in sample C compared to the pristine and sample B. The mechanism of such an abrupt change of current at the SET and RESET point is understood through the vacancy filamentary model. Now, the endurance of the device is examined by plotting the HRS and LRS with the increasing number of cycles, as shown in Fig. \ref{retain1}(b). The current was read at 0.3 V. The graph shows that HRS exhibits a negligible fluctuation with the number of cycles, whereas the LRS is relatively stable with almost zero slope. The clear ON-OFF is sustained up to the 100th cycle. The gradual increase in the gap between LRS and HRS (as HRS exhibits a positive slope) suggests that the device can store information for a long period of time. Hence, ion implantation with proper ion energy and species can be an alternate technique to improve the endurance capability of RRAM devices.

The schematic of the filamentary model through oxygen vacancy for understanding the current conduction mechanism in sample C is shown in Fig. \ref{5}. Lee et al. \cite{lee2010model} reported that lower formation energy is required for oxygen vacancy than Ni vacancy, and the certain configured oxygen vacancies form clusters help to form the conducting path. In sample C, high fluence boosts the sufficient number of vacancies in the matrix, which helps in filament formation. In zero bias condition (Fig. \ref{5}(a)), no vacancy filament formed, although a lot of vacancy defects are already present in the system. After applying the sufficient positive bias voltage, the vacancies and oxygen ions are generated. The oxygen ions will migrate toward the Ag electrode due to high drift velocity compared to vacancies. The existing large amount of induced and generated vacancies in the matrix accumulate to form the vacancy filament. The path between the top and bottom electrode is completed through the formation of vacancy filament at the SET point (Fig. \ref{5}(b)), and the current jumps abruptly to reach the LRS state. In the RESET process, the oxygen ions migrate back toward the bottom electrode. During migration, they recombine with the oxygen vacancies. The quick annihilation of the vacancies will rupture the filament (Fig. \ref{5}(c)), and the current will drop sharply at the RESET voltage. Hence, the formation and rupture of the vacancy filament explain the SET and RESET process of the ITO/NiO/Ag device (implanted with a sufficient high fluence of  2$\times$10$^{16}$ ions/cm$^2$). In neuromorphic systems, synaptic weights play a crucial role in determining the strength of the connections between neurons. Digital RRAM cells can be used to store these weights by encoding them in different resistance states. By adjusting the device to a SET or RESET state, RRAM can attain various resistance levels, allowing for precise modulation of synaptic weights. The quick SET and RESET operations of RRAM devices provide rapid updates to these weights, enabling high-speed learning and processing. The delayed SET with increasing cycles is compatible with slow learning, whereas the quick RESET indicates high processing. The analog switching is modulated by the interface-dominated Schottky effect at the NiO/Ag and ITO/NiO interface, whereas the bulk-dominated vacancy filamentary model influences the digital switching in sample C. The NiO film is modified by the creation of a lot of vacancy defects using Cu ion implantation, which plays the underlying role of transforming analog to digital switching of ITO/NiO/Ag memristor device after the threshold fluence.

\subsection{\label{sec:level4} CONCLUSION:}

In conclusion, we investigated the analog to digital transformation of bipolar resistive switching in 100 KeV Cu-implanted ITO/NiO/Ag memristor devices. Pristine and sample B show analog switching, but the digital switching characteristic developed in sample C. The Schottky barrier is formed at the NiO/Ag and ITO/NiO interfaces. The observed analog switching in pristine and sample B is interface barrier dominant due to the Schottky effect rather than the bulk domination. The reduction of $\phi_B$ for both the polarity increases the current conduction in pristine with the increase of the number of cycles. The opposite scenario, i.e., the lowering in current due to an increase of $\phi_B$ with the number of cycles, is observed in sample B. The combination of pristine and sample B can be utilized as a learning and forgetting process for neuromorphic applications. The sudden jump of current from OFF to ON and ON to OFF state at SET and RESET voltage in sample C is elucidated by the formation and rupturing of bulk-dominated vacancy filamentary model. In this case, the high Cu ion fluence boosts the vacancies in sample C above the critical vacancies (compared to pristine and sample B) which is sufficient in forming the filament for exhibiting such digital switching. Slow learning is associated with high SET points with the increasing number of cycles, whereas quick SET exhibits the possibility of high processing.

\begin{acknowledgments}

The authors thank the National Institute of Science Education and Research (NISER), DAE, Government of India, for funding this research work through project number RIN-4001. Sourav Bhakta and Pratap Sahoo also acknowledge the staff of the Inter-University Accelerator Center, New Delhi, for their constant support and for providing a stable beam for 100 KeV Cu ion implantation.

\end{acknowledgments}

\nocite{*}
\bibliography{aipsamp}

\end{document}